\documentclass[letterpaper]{article} % DO NOT CHANGE THIS
\usepackage{aaai24}  % DO NOT CHANGE THIS
\usepackage{times}  % DO NOT CHANGE THIS
\usepackage{helvet}  % DO NOT CHANGE THIS
\usepackage{courier}  % DO NOT CHANGE THIS
\usepackage[hyphens]{url}  % DO NOT CHANGE THIS
\usepackage{graphicx} % DO NOT CHANGE THIS
\def\UrlFont{\rm}  % DO NOT CHANGE THIS
\usepackage{natbib}  % DO NOT CHANGE THIS AND DO NOT ADD ANY OPTIONS TO IT
\usepackage{caption} % DO NOT CHANGE THIS AND DO NOT ADD ANY OPTIONS TO IT
\usepackage{algorithm}
\usepackage{algorithmic}
\usepackage{amsmath}
\usepackage{url}

\usepackage{newfloat}
\usepackage{listings}
\DeclareCaptionStyle{ruled}{labelfont=normalfont,labelsep=colon,strut=off} % DO NOT CHANGE THIS
\floatstyle{ruled}
\newfloat{listing}{tb}{lst}{}
\floatname{listing}{Listing}
\title{Concept-Guided LLM Agents for Human-AI Safety Codesign}
\author{
    Florian Geissler\textsuperscript{\rm 1},
    Karsten Roscher\textsuperscript{\rm 1},
    and Mario Trapp\textsuperscript{\rm 1, 2}
}
\affiliations {
    \textsuperscript{\rm 1}
    Fraunhofer IKS, Fraunhofer Institute for Cognitive Systems IKS, Munich, Germany \\  
    \textsuperscript{\rm 2}
    School of Computation, Information and Technology, Technical University of Munich, Germany \\
    \{florian.geissler, karsten.roscher, mario.trapp\}@iks.fraunhofer.de, mario.trapp@tum.de
}

\newcommand{\xml}{\textit{xml}}

\begin{document}

\maketitle

\begin{abstract}
Generative AI is increasingly important in software engineering, including safety engineering, where its use ensures that software does not cause harm to people. 
This also leads to high quality requirements for generative AI. Therefore, the simplistic use of Large Language Models (LLMs) alone will not meet these quality demands. It is crucial to develop more advanced and sophisticated approaches that can effectively address the complexities and safety concerns of software systems. Ultimately, humans must understand and take responsibility for the suggestions provided by generative AI to ensure system safety.
To this end, we present an efficient, hybrid strategy to leverage LLMs for safety analysis and Human-AI codesign.
In particular, we develop a customized LLM agent that uses elements of prompt engineering, heuristic reasoning, and retrieval-augmented generation to solve tasks associated with predefined safety concepts, in interaction with a system model graph. The reasoning is guided by a cascade of micro-decisions that help preserve structured information. We further suggest a graph verbalization which acts as an intermediate representation of the system model to facilitate LLM-graph interactions. Selected pairs of prompts and responses relevant for safety analytics illustrate our method for the use case of a simplified automated driving system.
\end{abstract}

\section{Introduction}
The advent of transformer-based \cite{Vaswani2017} large language models (LLMs) has sparked enormous popularity of generative artificial intelligence (AI) for creative, text-based tasks. Representative of this trend is the reported record of OpenAI's ChatPGT for the fastest growing user base of all times in February 2023 \cite{Reuters}.
The high quality of auto-generated text has inspired the exploration of LLMs for tasks that involve structured data, such as knowledge graphs. In particular, a widely desired use case is the application of LLMs to safety-analytical tasks \cite{Jin2023, Pan2023, Wang2023}, for example, reasoning about fault propagation in model graphs and associated risks. Attempts to make use of LLMs for a formal hazard analysis have been made with moderate success \cite{Diemert2023}.

A particular challenge lies in the transformation of verbalized content to a structured graph information, and vice versa. Since the text generation process is of statistical nature, thus not bound by logical constraints, LLM responses will not necessarily preserve information structures of a given input. Existing strategies to tackle this problem can be broadly categorized as follows \cite{Jin2023}: 1) Prompt-engineering to encourage specific output formats or structural rules; 2) Heuristic or algorithmic reasoning: Encouraging the LLM to perform chain-of-thought (CoT) reasoning, for example, to solve the problem step by step following self-generated or predetermined instructions; 3) Making use of external knowledge using retrieval-augmented generation (RAG), for example via non-AI tools; 4) Fine-tuning or retraining models to empirically minimize the loss of structured information, see e.g., GraphGPT \cite{Tang2023}.

\begin{figure}[t]
\centering
\includegraphics[width=0.45\textwidth]{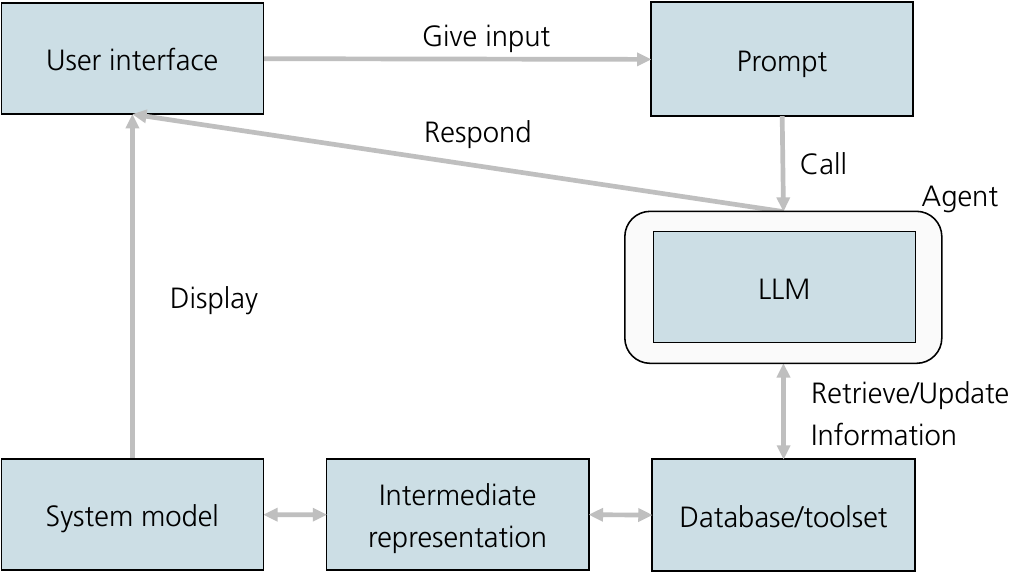} % Reduce the figure size so that it is slightly narrower than the column.
\caption{Layout of the Human-AI safety codesign framework: A user interacts with the LLM agent through a chat prompt and receives a text response. The LLM agent interfaces to a database containing a graph description of the system model in IR, as well as safety concepts, and analysis tools. System manipulations can update the database and alter the system model. The system model and its changes are displayed to the user.}
\label{fig_system}
\end{figure}

\begin{figure*}[htb!]
\centering
\includegraphics[width=0.9\textwidth]{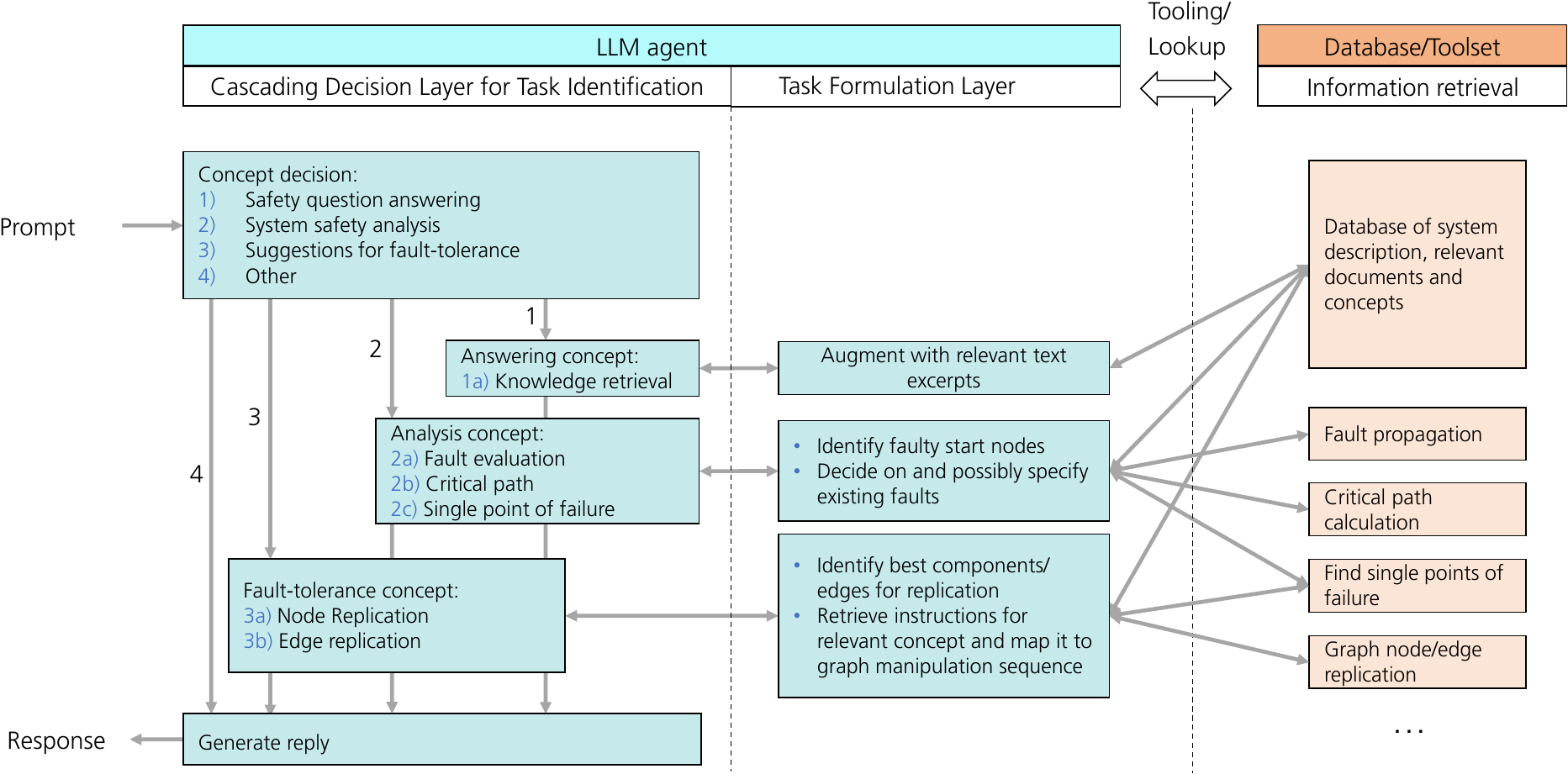} % Reduce the figure size so that it is slightly narrower than the column.
\caption{Outline of the workflow of LLM agent decisions and database interactions. The custom agent runs a cascading decision layer to identify the task type, and a subsequent layer to formulate the task for information retrieval. The latter looks up knowledge from a vectorstore database, or uses functional tools to calculate for example the critical path.}
\label{fig_tools}
\end{figure*}

This article presents a concept-guided approach aimed at enhancing the capabilities of LLMs for graph analysis and manipulation, particularly in the context of safety-relevant developments. To achieve this goal, we have integrated the strengths of LLMs with the rigorous standards of safety engineering models and analysis. Starting from a structured system model graph, we first establish a verbalized intermediate representation (IR) of the system to facilitate the comprehension of its components and relationships by the LLM. A custom LLM agent is designed who deploys a hybrid strategy featuring the techniques 1) to 3) above: The agent performs a series of LLM calls to categorize and formalize the task at hand according to predefined concepts. Subsequently, RAG techniques are leveraged to offload structural computations to external functions. We test our method for the example of a simplified automated driving architecture, and present selected experimental results for the tasks of \textit{fault propagation}, \textit{critical path finding}, \textit{single-points-of-failure detection}, and \textit{node replication}.
Our work provides the basis for an interactive, LLM-based Human-AI safety codesign framework.

\section{Model}
The overall system layout is shown in Fig.~\ref{fig_system}. Key components of our architecture are explained in the following.

\textbf{System model and intermediate representation:} 
% Architecture analysis and design language (AADL)
% We start from the system model of a realistic use case. 
The system model contains the system architecture, including the system components and their interactions, as well as additional safety-related information required to model fault propagation within the system. To ensure interoperability with established industrial meta model formats, we create the system model with the Eclipse-based OSATE tool \cite{CarnegieMellonUniversity2023} and export the resulting \textit{ECore} file as \xml.
The elements of the \xml\ model (EClass, EReference, EAttribute) can be directly mapped onto the elements of a generic graph (nodes, edges, attributes). Even though LLMs are capable of reading and interpreting \xml\ structures directly, we find that inaccuracies can be reduced when operating with a system description that has a closer resemblance with natural language. Therefore, we further verbalize the \xml\ system model to an IR which takes an intuitive list structure:
\begin{lstlisting}[numbers=none]
Nodes:
    - Node 1
    - Node 2
    - ...
Edges:
    - Node 1 --> Node 2
    - ...
Attributes:
    - Node 1: Attribute 1
    - ...
\end{lstlisting}

This IR represents the basis of system-relevant information for the LLM agent. For safety analysis, we assume that each node is subject to possible failure.
In order to model the propagation of system faults, we populate graph node attributes with verbalized logic of fault gates (e.g., \textit{AND}, \textit{OR}, \textit{N-out-of-M}) to represent fault trees \cite{Avizienis2004, Trapp2016}. Further, start and end nodes of the system graph are specified with corresponding attributes. For the current proof-of-concept, we used an integrated system-safety model. However, in model-based safety engineering, long-lasting work has been conducted on integrating system and safety models \cite{DomisTrapp2008}, which can be used as a scalable basis for further development.

\textbf{LLM:} At the core, we use OpenAI's $GPT 3.5-turbo$ \cite{OpenAI} model for LLM inferences. As our concept-guided approach requires the LLM to solve a series of rather simple micro-decisions, we expect to see in further work, that also much smaller and simpler models such as LLama2 \cite{Touvron2023} or Mistral-7B \cite{Jiang2023} suffice for this purpose.

\begin{figure*}[hbt!]
\centering
\includegraphics[width=0.95\textwidth]{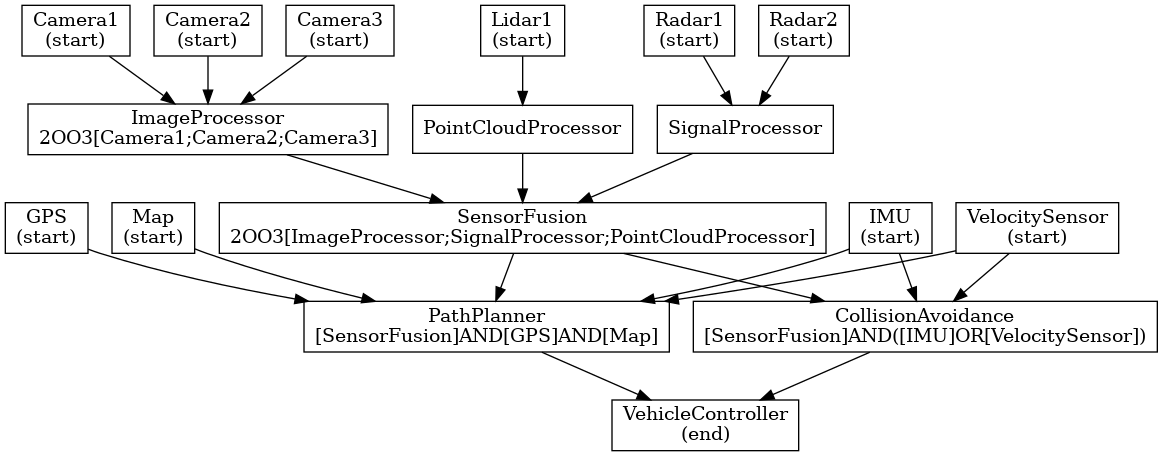} % Reduce the figure size so that it is slightly narrower than the column.
\caption{Example use case of a simplified automated driving system. The node labels denote the component name (top row) and the fault gate attribute below (if none is given, an AND fault gate of all inputs is assumed). $2OO3$ means that two out of three of the listed inputs are required. Start and end nodes are further labeled explicitly. The figure is a \textit{pydot} visualization of the \textit{ECore} file.}
\label{fig_usecase}
\end{figure*}

\textbf{LLM agent:}
LLMs are conveniently orchestrated by functional wrappers called \textit{agents}, which can be configured to self-induce an iterative chain of thoughts, or to interface automatically to external tools and sources of information. We here use the \textit{LangChain} library \cite{HarrisonChase} to design a customized agent. Importantly, we find that equipping an agent with multiple tools can quickly lead to inaccurate tool usage, unless the verbalized trigger conditions are well-separated in linguistic space. At the same time, for self-induced CoT flows, it is challenging to incorporate and assert procedural safety guardrails.

We therefore implement a different workflow, as illustrated in Fig.~\ref{fig_tools}: The agent cascades the input prompt through a network of micro-decisions, where a single LLM call associates the input with the best match among only $2$-$4$ predefined concepts in each decision. For example, as shown in Fig.~\ref{fig_tools}, in the first decision node the agent associates the input with one of four possible task concepts of \textit{Safety question answering}, \textit{System safety analysis}, \textit{Suggestions for fault-tolerance}, and \textit{Other}. To improve the quality of the micro-decision, we use few-shot learning \cite{Brown2020} by augmenting the prompts with around ten hand-crafted example triggers for each concept.
The last layer of our decision network is a \textit{Task formulation layer}, where the fully identified task concept is used together with the given prompt to refine a specific task prompt. Finally, the external tools or information lookup mechanisms that are provided for this specific task are called and the result is rephrased as a user response. We find that this method overcomes some of the challenges mentioned above: The CoT flow is tightly guided by predefined concepts, yet allows sufficient flexibility in the prompt formulation, and tool triggers cannot overlap since they are strictly associated with specific task concepts.

\textbf{Database/toolset:}
We provide a collection of information that is retrieved from a knowledge base, or generated on demand with deterministic functions. This includes, but is not limited to the following: The IR of the system, documentation about safety practices, a function to calculate fault propagation using the fault gates retrieved from the system IR, a function to calculate the critical path using Dijkstra's algorithm \cite{Dijkstra1959} for the shortest path and start and end nodes from the IR, a function to find the single points of failure, a function to replicate a given node or edge in the model IR. The usage of these tools is triggered in a predefined way depending on the identified task category.

\begin{table*}[hbt!]
\centering
\begin{tabular}{|p{2.5cm}|p{14cm}|}
    \hline
    Prompt & Response \\
    \hline
    \hline
    What happens if Radar1, Radar2 and IMU have a fault? & 
   {In this scenario, the faulty components are IMU, Radar1, Radar2, and SignalProcessor. When these components have a fault, it can have several consequences for the system. The IMU (Inertial Measurement Unit) is responsible for measuring the orientation, velocity, and acceleration of the vehicle. If the IMU has a fault, it can lead to inaccurate measurements, which can affect the overall navigation and control of the vehicle. The radars (Radar1 and Radar2) are used for detecting and tracking objects in the environment. If these radars have a fault, it can result in missed or incorrect detections, which can impact the perception and decision-making capabilities of the system. The SignalProcessor is responsible for processing the sensor data and extracting relevant information. If the SignalProcessor has a fault, it can lead to incorrect or incomplete data processing, which can affect the overall performance and reliability of the system. Overall, when these components have a fault, it can lead to degraded performance, reduced safety, and potentially compromised functionality of the system.} \\
    \hline
    % What happens if Lidar1 and VelocitySensor break? & 
    % {When Lidar1 and VelocitySensor break, the affected components in the system are VelocitySensor, Lidar1, and PointCloudProcessor. The fault in these components can have various consequences to the system. The faulty VelocitySensor can result in inaccurate measurement of the vehicle's velocity, which can affect the control and navigation of the system. The faulty Lidar1 can lead to a loss of reliable distance and object detection information, which can impact the perception and situational awareness of the system. The faulty PointCloudProcessor can result in incorrect processing and interpretation of the lidar data, which can affect the accuracy of object detection and tracking. It is important to address these faults and ensure proper functioning of the components to maintain the reliability and effectiveness of the system.} \\
    % \hline
    Show me the critical path. &
    {The critical path in the given scenario includes the following components: Camera1, Camera2, CollisionAvoidance, GPS, IMU, ImageProcessor, Lidar1, Map, PathPlanner, PointCloudProcessor, SensorFusion, and VehicleController. The critical path represents the sequence of components that are essential for the system's operation and have the longest duration in terms of time. These components are crucial for tasks such as perception, sensor fusion, planning, and control, and any faults or failures in these components can significantly impact the overall performance and safety of the system.} \\
    \hline
    Explain the critical path, given the last fault. &
    {The critical path in this scenario, considering the last reported faulty system components, consists of the following components: Camera1, Camera3, CollisionAvoidance, GPS, ImageProcessor, Lidar1, Map, PathPlanner, PointCloudProcessor, SensorFusion, VehicleController, and VelocitySensor. The critical path represents the sequence of components and processes that are essential for the system's operation and performance. Any delay or failure in these components can significantly impact the overall functionality and reliability of the system. It is important to note that the previously reported system faults, including IMU, Radar1, Radar2, and SignalProcessor, can also affect the critical path and further exacerbate the potential consequences to the system.} \\
    \hline
    What are the single points of failure? &
    {The single point of failure components in the given scenario are PathPlanner, VehicleController, Map, SensorFusion, CollisionAvoidance, and GPS. These components are considered single points of failure because if any of them were to fail, it could result in a complete system failure or significant degradation in the system's performance. For example, if the PathPlanner fails, the system may not be able to generate safe and efficient trajectories for the vehicle. Similarly, if the VehicleController fails, the system may lose control over the vehicle's movements. The Map, SensorFusion, CollisionAvoidance, and GPS components are also critical for accurate perception, situational awareness, and navigation, and their failure can have serious consequences for the overall system functionality.} \\
    \hline
    How would you make my system safer? &
    {To make your system safer, I would suggest performing replication on the component SensorFusion. SensorFusion should be replicated because it is a critical component that relies on multiple inputs from ImageProcessor, SignalProcessor, and PointCloudProcessor. If SensorFusion fails, it can lead to incorrect data being sent to PathPlanner and CollisionAvoidance, compromising the overall safety of the system. By replicating SensorFusion, the system can continue to function even if one instance of SensorFusion fails, ensuring that accurate and reliable data is provided to the other components.}\\
    \hline
\end{tabular}
\caption{Selected example prompts and responses. Refer to Fig.~\ref{fig_usecase} for the system description. The example in row three references the agent's memory, as \textit{last fault} refers to the last mentioned system fault in the chat history.}
\label{tab_results}
\end{table*}

\section{Experiments}
We test our model with the use case of a simplified automated driving system modeled in OSATE, see Fig.~\ref{fig_usecase}, here visualized with \textit{pydot} \cite{Carrera2021}.
Fault gates are implemented as respective node attributes and given in the figure below the node label, for example, the \textit{ImageProcessor} node will turn faulty if at least two out of three ($2OO3$) camera input nodes are faulty. 
%If no fault gate is given, a default \textit{AND} gate of all inputs is assumed.
To verify the effectiveness of our concept-guided agent, we test our approach with sample questions in the domain of system safety design. Representative examples and results are given in Tab.~\ref{tab_results}. In all examples, we find that the tasks have been accurately identified and formulated, and the correct information was retrieved with the appropriate tools. For example, the agent can make suggestions about modifying the graph to improve fault tolerance. In our implementation, this is based on the predefined concept of redundancy and the usage of the tool which finds single points of failures to select the best replication candidate. To modify the system graph, the agent uses the graph replication tool and updates the \xml\ model accordingly.
We further note that, in the response, the agent leverages general knowledge to relate the functionality of the relevant components with its impact on safety.

\section{Conclusion and Outlook}
Our approach of a concept-guided LLM agent overcomes two challenges typically encountered when solving generative text tasks on structured data: 1) Chain-of-thought processes derail and no longer comply with a set of required rules, or miss their goals, 2) Trigger conditions to retrieve external information via tools become inaccurate when the number of possible options increases. Instead, we here guide the thought process with a series of micro-decisions to ensure that the right tool triggers are met. 
We envision to extend our setup to more concepts and larger decision cascades, such that more complex tasks can be tackled. 
Our work provides the basis for an interactive framework where LLMs assist in Human-AI safety codesign.

\section{Acknowledgments}
This work was funded by the Bavarian Ministry for Economic Affairs, Regional Development and Energy as part of a project to support the thematic development of the Institute for Cognitive Systems.

% \begin{quote}
% \begin{small}
\bibliography{Bib2}

\begin{thebibliography}{18}
\providecommand{\natexlab}[1]{#1}

\bibitem[{Avi{\v{z}}ienis et~al.(2004)Avi{\v{z}}ienis, Laprie, Randell, and
  Landwehr}]{Avizienis2004}
Avi{\v{z}}ienis, A.; Laprie, J.~C.; Randell, B.; and Landwehr, C. 2004.
\newblock {Basic concepts and taxonomy of dependable and secure computing}.
\newblock \emph{IEEE Transactions on Dependable and Secure Computing}, 1(1):
  11--33.

\bibitem[{Brown et~al.(2020)Brown, Mann, Ryder, Subbiah, Kaplan, Dhariwal,
  Neelakantan, Shyam, Sastry, Askell, Agarwal, Herbert-Voss, Krueger, Henighan,
  Child, Ramesh, Ziegler, Wu, Winter, Hesse, Chen, Sigler, Litwin, Gray, Chess,
  Clark, Berner, McCandlish, Radford, Sutskever, and Amodei}]{Brown2020}
Brown, T.~B.; Mann, B.; Ryder, N.; Subbiah, M.; Kaplan, J.; Dhariwal, P.;
  Neelakantan, A.; Shyam, P.; Sastry, G.; Askell, A.; Agarwal, S.;
  Herbert-Voss, A.; Krueger, G.; Henighan, T.; Child, R.; Ramesh, A.; Ziegler,
  D.~M.; Wu, J.; Winter, C.; Hesse, C.; Chen, M.; Sigler, E.; Litwin, M.; Gray,
  S.; Chess, B.; Clark, J.; Berner, C.; McCandlish, S.; Radford, A.; Sutskever,
  I.; and Amodei, D. 2020.
\newblock {Language models are few-shot learners}.
\newblock \emph{Advances in Neural Information Processing Systems}, 2020-Decem.

\bibitem[{{Carnegie Mellon University}(2023)}]{CarnegieMellonUniversity2023}
{Carnegie Mellon University}. 2023.
\newblock {OSATE 2.13}.
\newblock https://osate.org/.
\newblock Accessed: 2023-12-01.

\bibitem[{Carrera(2021)}]{Carrera2021}
Carrera, E. 2021.
\newblock pydot.
\newblock https://pypi.org/project/pydot/.
\newblock Accessed: 2023-12-01.

\bibitem[{Diemert and Weber(2023)}]{Diemert2023}
Diemert, S.; and Weber, J.~H. 2023.
\newblock {Can Large Language Models Assist in Hazard Analysis?}
\newblock In \emph{Lecture Notes in Computer Science (including subseries
  Lecture Notes in Artificial Intelligence and Lecture Notes in
  Bioinformatics)}, volume 14182 LNCS, 410--422.
\newblock ISBN 9783031409523.

\bibitem[{Dijkstra(1959)}]{Dijkstra1959}
Dijkstra, E.~W. 1959.
\newblock {A note on two problems in connexion with graphs}.
\newblock \emph{Numer. Math.}, 271: 269--271.

\bibitem[{Domis and Trapp(2008)}]{DomisTrapp2008}
Domis, D.; and Trapp, M. 2008.
\newblock Integrating Safety Analyses and Component-Based Design.
\newblock In Harrison, M.~D.; and Sujan, M.-A., eds., \emph{Computer Safety,
  Reliability, and Security}, 58--71. Berlin, Heidelberg: Springer Berlin
  Heidelberg.
\newblock ISBN 978-3-540-87698-4.

\bibitem[{{Harrison Chase}(2022)}]{HarrisonChase}
{Harrison Chase}. 2022.
\newblock {LangChain}.
\newblock https://github.com/langchain-ai/langchain.
\newblock Accessed: 2023-12-01.

\bibitem[{Jiang et~al.(2023)Jiang, Sablayrolles, Mensch, Bamford, Chaplot,
  de~las Casas, Bressand, Lengyel, Lample, Saulnier, Lavaud, Lachaux, Stock,
  Scao, Lavril, Wang, Lacroix, and Sayed}]{Jiang2023}
Jiang, A.~Q.; Sablayrolles, A.; Mensch, A.; Bamford, C.; Chaplot, D.~S.; de~las
  Casas, D.; Bressand, F.; Lengyel, G.; Lample, G.; Saulnier, L.; Lavaud,
  L.~R.; Lachaux, M.-A.; Stock, P.; Scao, T.~L.; Lavril, T.; Wang, T.; Lacroix,
  T.; and Sayed, W.~E. 2023.
\newblock Mistral 7B.
\newblock arXiv:2310.06825.

\bibitem[{Jin et~al.(2023)Jin, Liu, Han, Jiang, Ji, and Han}]{Jin2023}
Jin, B.; Liu, G.; Han, C.; Jiang, M.; Ji, H.; and Han, J. 2023.
\newblock Large Language Models on Graphs: A Comprehensive Survey.
\newblock arXiv:2312.02783.

\bibitem[{OpenAI(2023)}]{OpenAI}
OpenAI. 2023.
\newblock {ChatGPT 3.5-turbo}.
\newblock https://openai.com/.
\newblock Accessed: 2023-12-01.

\bibitem[{Pan et~al.(2023)Pan, Luo, Wang, Chen, Wang, and Wu}]{Pan2023}
Pan, S.; Luo, L.; Wang, Y.; Chen, C.; Wang, J.; and Wu, X. 2023.
\newblock Unifying Large Language Models and Knowledge Graphs: A Roadmap.
\newblock arXiv:2306.08302.

\bibitem[{Reuters(2023)}]{Reuters}
Reuters. 2023.
\newblock {ChatGPT sets record for fastest-growing user base - analyst note}.
\newblock
  https://www.reuters.com/technology/chatgpt-sets-record-fastest-growing-user-base-analyst-note-2023-02-01/.
\newblock Accessed: 2023-12-01.

\bibitem[{Tang et~al.(2023)Tang, Yang, Wei, Shi, Su, Cheng, Yin, and
  Huang}]{Tang2023}
Tang, J.; Yang, Y.; Wei, W.; Shi, L.; Su, L.; Cheng, S.; Yin, D.; and Huang, C.
  2023.
\newblock GraphGPT: Graph Instruction Tuning for Large Language Models.
\newblock arXiv:2310.13023.

\bibitem[{Touvron et~al.(2023)Touvron, Martin, Stone, Albert, Almahairi,
  Babaei, Bashlykov, Batra, Bhargava, Bhosale, Bikel, Blecher, Ferrer, Chen,
  Cucurull, Esiobu, Fernandes, Fu, Fu, Fuller, Gao, Goswami, Goyal, Hartshorn,
  Hosseini, Hou, Inan, Kardas, Kerkez, Khabsa, Kloumann, Korenev, Koura,
  Lachaux, Lavril, Lee, Liskovich, Lu, Mao, Martinet, Mihaylov, Mishra,
  Molybog, Nie, Poulton, Reizenstein, Rungta, Saladi, Schelten, Silva, Smith,
  Subramanian, Tan, Tang, Taylor, Williams, Kuan, Xu, Yan, Zarov, Zhang, Fan,
  Kambadur, Narang, Rodriguez, Stojnic, Edunov, and Scialom}]{Touvron2023}
Touvron, H.; Martin, L.; Stone, K.; Albert, P.; Almahairi, A.; Babaei, Y.;
  Bashlykov, N.; Batra, S.; Bhargava, P.; Bhosale, S.; Bikel, D.; Blecher, L.;
  Ferrer, C.~C.; Chen, M.; Cucurull, G.; Esiobu, D.; Fernandes, J.; Fu, J.; Fu,
  W.; Fuller, B.; Gao, C.; Goswami, V.; Goyal, N.; Hartshorn, A.; Hosseini, S.;
  Hou, R.; Inan, H.; Kardas, M.; Kerkez, V.; Khabsa, M.; Kloumann, I.; Korenev,
  A.; Koura, P.~S.; Lachaux, M.-A.; Lavril, T.; Lee, J.; Liskovich, D.; Lu, Y.;
  Mao, Y.; Martinet, X.; Mihaylov, T.; Mishra, P.; Molybog, I.; Nie, Y.;
  Poulton, A.; Reizenstein, J.; Rungta, R.; Saladi, K.; Schelten, A.; Silva,
  R.; Smith, E.~M.; Subramanian, R.; Tan, X.~E.; Tang, B.; Taylor, R.;
  Williams, A.; Kuan, J.~X.; Xu, P.; Yan, Z.; Zarov, I.; Zhang, Y.; Fan, A.;
  Kambadur, M.; Narang, S.; Rodriguez, A.; Stojnic, R.; Edunov, S.; and
  Scialom, T. 2023.
\newblock Llama 2: Open Foundation and Fine-Tuned Chat Models.
\newblock arXiv:2307.09288.

\bibitem[{Trapp(2016)}]{Trapp2016}
Trapp, M. 2016.
\newblock {Assuring Functional Safety in Open Systems of Systems}.
\newblock https://nbn-resolving.de/urn:nbn:de:hbz:386-kluedo-44221.

\bibitem[{Vaswani et~al.(2017)Vaswani, Shazeer, Parmar, Uszkoreit, Jones,
  Gomez, Kaiser, and Polosukhin}]{Vaswani2017}
Vaswani, A.; Shazeer, N.; Parmar, N.; Uszkoreit, J.; Jones, L.; Gomez, A.~N.;
  Kaiser, {\L}.; and Polosukhin, I. 2017.
\newblock {Attention is all you need}.
\newblock \emph{Advances in Neural Information Processing Systems}, 2017-Decem:
  5999--6009.

\bibitem[{Wang et~al.(2023)Wang, Feng, He, Tan, Han, and Tsvetkov}]{Wang2023}
Wang, H.; Feng, S.; He, T.; Tan, Z.; Han, X.; and Tsvetkov, Y. 2023.
\newblock Can Language Models Solve Graph Problems in Natural Language?
\newblock arXiv:2305.10037.

\end{thebibliography}
% \end{small}
% \end{quote}

\end{document}